\documentclass[journal]{IEEEtran}

\usepackage{graphicx}
\usepackage{amsmath}
\usepackage{amsfonts}
\usepackage{verbatim}
\usepackage{color}
\usepackage{ulem}

\usepackage[hypertex]{hyperref}

\newcommand{\blue} {\ensuremath{^1\!S_0\,-\,^1\!P_1}}

\begin{document}

\title{Minimizing the Dick Effect in an Optical Lattice Clock}

%\author{Philip G. Westergaard, J\'er\^ome Lodewyck and Pierre
%Lemonde} \affiliation{LNE-SYRTE, Observatoire de Paris, CNRS, UPMC
%\\ 61 avenue de l'Observatoire, 75014 Paris, France}

\author{\authorblockN{Philip G. Westergaard, J\'er\^ome Lodewyck and Pierre Lemonde}
\authorblockA{LNE-SYRTE, Observatoire de Paris, CNRS, UPMC \\ 61
avenue de l'Observatoire, 75014 Paris, France}}

\maketitle

\begin{abstract}
We discuss the minimization of the Dick effect in an optical
lattice clock. We show that optimizing the time sequence of
operation of the clock can lead to a significant reduction of the
clock stability degradation by the frequency noise of the
interrogation laser. By using a non-destructive detection of the
atoms, we are able to recycle most of the atoms between cycles and
consequently to strongly reduce the time spent capturing the atoms
in each cycle. With optimized parameters, we expect a fractional
Allan deviation better than $2 \cdot 10^{-16}\,\tau^{-1/2}$ for
the lattice clock.
\end{abstract}

\section{Introduction}
Combined with a superb frequency accuracy, superior ultimate
stabilities have been advocated as appealing advantages of optical
lattice clocks\,\cite{Katopal03}. In such devices, optical
resonances with linewidth down to 2\,Hz have been
observed\,\cite{Boyd06}. For a typical atom number of $10^5$ the
corresponding standard quantum limit of the clock Allan deviation
lies below $10^{-17}\,\tau^{-1/2}$ with $\tau$ being the averaging
time in seconds. While vast improvements have been performed over
the last few
years\,\cite{Hong08,opticalpumping,Ludlow08,Barber08}, the
stability of actual lattice clocks is presently more than two
orders of magnitude above this ``Holy Grail''. One stumbles upon
the Dick effect, by which the probe laser frequency noise is
converted down to low Fourier frequencies by the sampling process
inherent to the clock's cyclic
operation\,\cite{Dick87,Santarelli98,Quessada03}. A strenuous
effort is presently going on to further reduce the noise of
ultra-stable laser
sources\,\cite{Young99,Nazarova06,Ludlow07,Webster07,Millo09} but
quite hard limitations like the thermal noise of high finesse
Fabry-P\'{e}rot cavities limit progress in that
direction\,\cite{Numata04}. Comparatively little effort has been
put so far on the optimization of the time sequence for the
operation of lattice clocks for reducing the Dick effect. We show
here that following this direction can lead to very significant
improvements.

A key parameter for the Dick effect is the dead time of the clock
cycle, during which atoms are prepared (captured, cooled,
optically pumped) and detected and do not experience the probe
laser frequency noise. This loss of information leads to the
frequency stability degradation. In order to dwarf the dead time
of the experiment, we propose to keep the atoms from one clock
cycle to the next which is made possible by a non-destructive
measurement scheme\,\cite{Lodewyck09}. We discuss here in detail
the potential gain in terms of frequency stability that can be
achieved using this detection scheme.

In section\,\ref{sec:Dick}, we give a quantitative discussion of
the Dick effect in the limit where the dead time approaches 0. We
show that for dead times below 100\,ms, the limitation of the
Allan deviation due to the Dick effect can be reduced to below
$10^{-16}\,\tau^{-1/2}$ using Ramsey spectroscopy and
state-of-the-art ultra-stable lasers. In
section\,\ref{sec:Non-destr} the new non-destructive detection
scheme is described. Finally, section\,\ref{sec:SrStability}
discusses the optimization of a Sr lattice clock sequence using
the non-destructive scheme, and gives an estimate on the expected
stability of the clock.

\section{The Dick Effect in the Low Dead Time Limit\label{sec:Dick}}

In a sequentially operated atomic clock, the response of the atoms
to the interrogation oscillator frequency fluctuations $\delta
\omega (t)$ is dictated by the sensitivity function $g(t)$. The
change in transition probability $\delta P$ due to frequency noise
is given by
\begin{equation}
    \delta P = \frac{1}{2} \int {  g(t) \delta \omega (t) dt },   \label{eq:SensFunc}
\end{equation}
where the integral is taken over one clock cycle. The appearance
of $g(t)$ depends on the type of interrogation used. In an optical
lattice clock either Rabi or Ramsey interrogation can be used. We
call $T_i$ the duration of the interrogation $\pi$-pulse in the
Rabi case and $\tau_p$ the duration of each of the two
$\pi/2$-pulses and $T$ the free evolution time in the Ramsey case.
We define the duty cycle $d= \frac{T_i}{T_c}$ (Rabi) and
$d=\frac{2\tau_p + T}{T_c}$ (Ramsey) with $T_c$ being the duration
of the clock cycle.

\begin{figure}
\begin{center}
    \includegraphics[width=1.0\columnwidth]{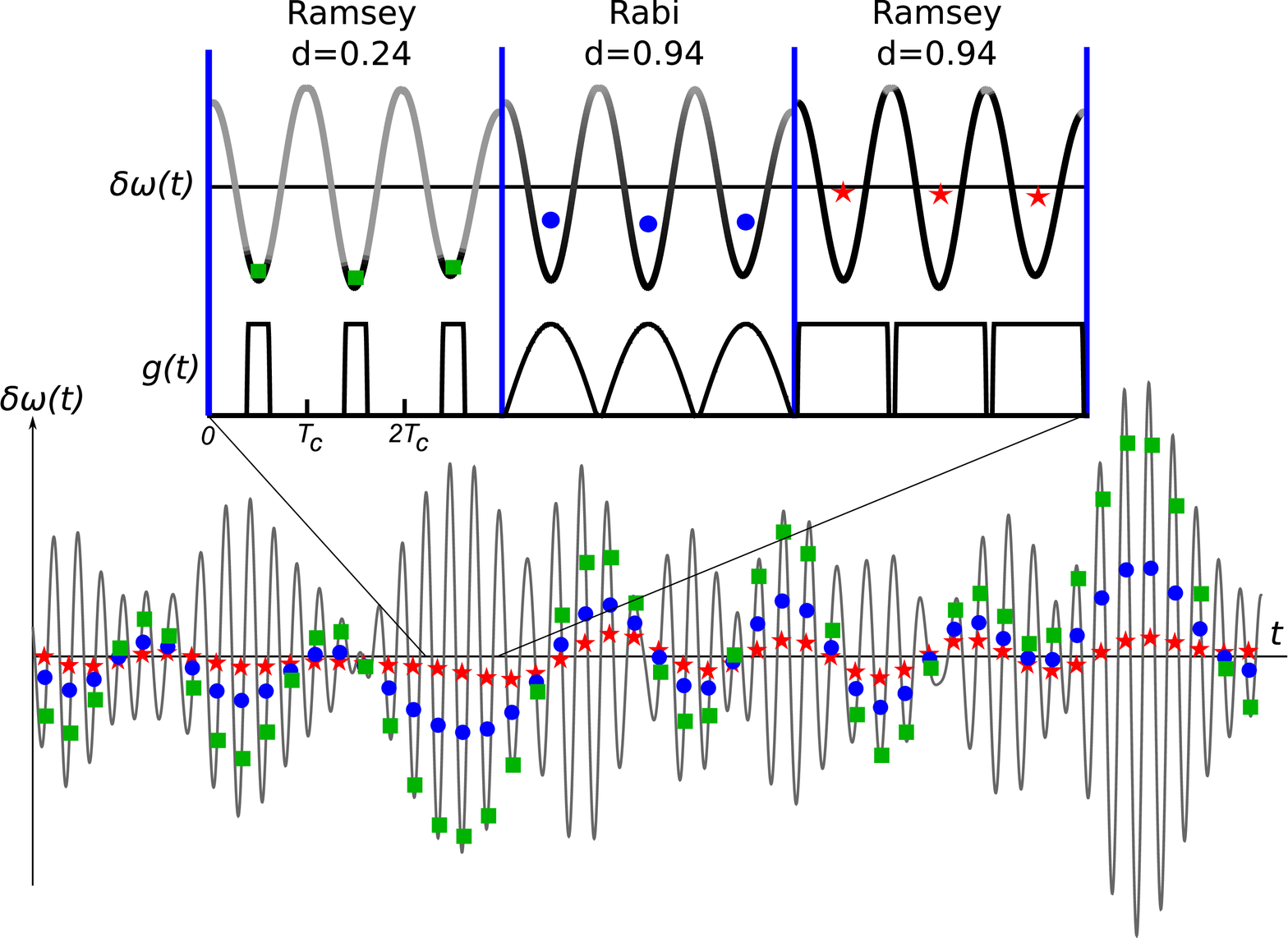}
    \caption{Simulated frequency noise $\delta \omega(t)$ of the interrogation oscillator filtered
    around the cycle frequency $f_c=1/T_c$ with a bandwidth of $0.3\,f_c$.
    The points show the weighted average $\int g(t) \delta \omega(t) dt/\int g(t)
    dt$ for Rabi interrogation with duty cycle $d=0.94$ (circles) and for Ramsey interrogation
     with duty cycles $d = 0.24$ (squares) and $d=0.94$ (stars).
    The inset shows how $\delta \omega(t)$ is sampled over 3 cycles for the three different sensitivity functions $g(t)$.
    }
    \label{fig:NoiseSampling}
\end{center}
\end{figure}

\begin{figure}
\begin{center}
    \includegraphics[width=0.8\columnwidth]{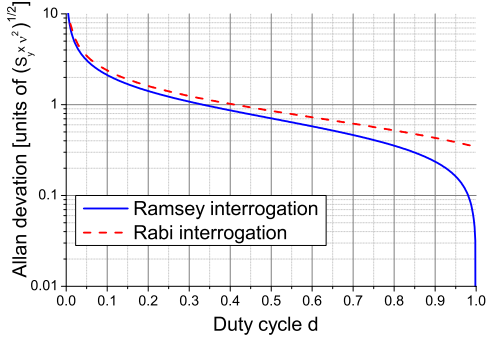}
    \caption{Dick limited Allan deviation $\sigma_y (\tau=1\textrm{ s})$ for white
     frequency noise for Rabi and Ramsey interrogation.
     In the case of Ramsey interaction, the Dick effect vanishes for $d \rightarrow 1$ if $\tau_p$ is kept smaller than $T_d$.
     The curves are computed using \eqref{eq:AllanVar}.}
    \label{fig:RabiRamsey}
\end{center}
\end{figure}

Fig. \ref{fig:NoiseSampling} gives a clear graphic illustration of
the Dick effect and of the role of the dead time. The figure shows
numerically generated noise around the cycle frequency $f_c=1/T_c$
with a bandwidth of $0.3\,f_c$. The noise of the oscillator enters
in the clock measurement as the time average of $ \delta
\omega(t)$ weighted by $g(t)$, according to Eq. \eqref{eq:SensFunc}.
For a small duty cycle (squares in Fig. \ref{fig:NoiseSampling})
only the maxima of the relevant noise components contribute to the
measurement, resulting in a large dispersion of the measured
frequency. When the duty cycle $d$ approaches 1, the sensitivity
function comprises almost the totality of each cycle, and the
frequency fluctuations of the interrogation oscillator are
averaged out. This averaging effect is almost perfect in the case
of Ramsey interaction (stars in Fig.\,\ref{fig:NoiseSampling} for
$d=0.94$) since the sensitivity function is a constant during the
free evolution period. As the dead time $T_d$ used to prepare and
detect atoms approaches 0, the measurement noise totally vanishes
provided the interrogation pulses are kept short enough
($\tau_p\ll T_d$). The situation is quite different for Rabi
interrogation (circles in Fig.\,\ref{fig:NoiseSampling}), since
the sinusoidal shape of $g(t)$ enfeebles the efficiency of the
averaging process. The averaging effect and the different behavior
depending on the interrogation scheme is further illustrated in
Fig. \ref{fig:RabiRamsey}, where the Allan deviation as a function
of the duty cycle is plotted\,\footnote{For illustration purposes
we chose to plot Fig. \ref{fig:RabiRamsey} in the case where the
interrogation laser exhibits white frequency noise. This is the
only type of noise where the Allan deviation only depends on the
duty cycle, and not on the specific parameters chosen. For
subsequent discussion, however, we will assume a more
experimentally realistic flicker frequency noise.}. Due to this
clear advantage of Ramsey interrogation, we restrict further
analysis to this case only.

The limitation of the fractional Allan variance due to the
interrogation laser frequency noise is given by
\cite{Santarelli98}
\begin{equation}
    \sigma_y^2(\tau) = \frac{1}{\tau g_0^2} \sum_{m=1}^\infty { \left| g_{m} \right|^2 S_y(m/T_c) },   \label{eq:AllanVar}
\end{equation}
where $S_y(f)$ is
the one-sided power spectral density of the relative frequency
fluctuations of the free running interrogation oscillator taken at
Fourier frequencies $m/T_c$. The Fourier coefficients of $g(t)$ are given by
\begin{eqnarray}
      g_{m}  &=& \frac{1}{T_c} \int_0^{T_c} g(t) e^{-2\pi i m t/T_c}
      dt.
           \label{eq:FourierComps}
\end{eqnarray}

State-of-the-art interrogation laser stabilization is performed by
locking the laser frequency to an ultra-stable Fabry-Perot cavity.
In the following, we assume that the dominant source of noise is
the thermal noise of the cavity $S_y(f) = h_{-1}f^{-1}/\nu^2$ with
$\nu$ being the clock frequency ($\nu=4.29 \cdot 10^{14} \,$Hz for
a Sr lattice clock). We take $h_{-1}=4 \cdot 10^{-2}$ Hz$^4$ which
is a worst case estimate for the ULE cavity with fused silica
mirrors described in \cite{Millo09}. It corresponds to a constant Allan
standard deviation of $6\cdot 10^{-16}$.

\begin{figure}
\begin{center}
    \includegraphics[width=1.0\columnwidth]{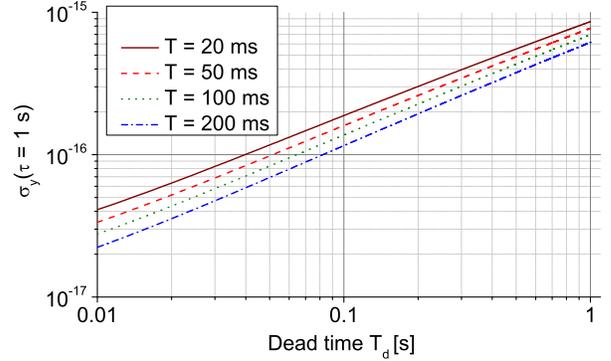}
    \caption{Fractional Allan deviation $\sigma_y(\tau=1\, \textrm{s})$ {\it vs} dead time for various durations of the Ramsey interrogation $T$. The interrogation laser noise is supposed to be flicker dominated (see text). The duty cycle is $d=0.02$ for $T_d=1.0\,$s and $T=0.02\,$s, and $d=0.95$ for $T_d=0.01\,$s and $T=0.2 \,$s.}
    \label{fig:FixedRatio}
\end{center}
\end{figure}

Figure \ref{fig:FixedRatio} displays the Allan deviation computed
numerically using \,\eqref{eq:AllanVar} as a function of dead time
$T_d$ for various $T$. We choose $\tau_p=5\,$ms which is
significantly shorter than the shortest $T_d$ considered here and
still long enough to keep the $\tau_p$ dependent frequency shifts
(light shift, line pulling by other atomic resonances, etc...)
reasonably small (see \cite{Taichenachev09} and references
therein). Fig. \ref{fig:FixedRatio} is another illustration of the
averaging process discussed above. In present optical lattice
clocks, the dead time is on the order of 1\,s and the limitation
of the clock stability due to the Dick effect is
close\footnote{Measured Allan deviations are somehow higher than
the value calculated here (experimental state-of-the-art value is
about $2\cdot10^{-15}$ for one second \cite{calcium}). This
results mainly from the fact that the interrogation lasers used
for these experiments are referenced to cavities with ULE mirror
substrates, which exhibit substantially higher thermal noise than
the cavities considered here.} to $10^{-15}$. Reducing this dead
time down to 10\,ms would improve the clock stability by almost
two orders of magnitude. This consideration motivated the
development of the non-destructive detection scheme which is
presented in the next section. Note also that for a given dead
time, it is desirable to lengthen as much as possible the Ramsey
interaction. This is true as long as the linear model giving
Eq. \eqref{eq:SensFunc} and \eqref{eq:AllanVar} holds, {\it i.e.} as long as the
interrogation laser frequency fluctuations remain much smaller
than the width of the Ramsey fringes. With the level of noise
chosen for plotting Fig.\,\ref{fig:FixedRatio} - that is,
frequency fluctuations of the interrogation oscillator on the
order of $0.3 \,$Hz - the model therefore holds for Ramsey times
up to about 200\,ms.

\section{Non-Destructive Measurement \label{sec:Non-destr}}

We briefly recall here the main features of the detection scheme
which allows optimization of the clock stability as discussed in
section\,\ref{sec:SrStability}. More details can be found in
\cite{Lodewyck09}.

\subsection{Experimental Setup \label{section:setup}}
The scheme is based on the measurement of the phase shift
accumulated by a weak probe beam tuned close to an atomic
resonance when passing through the atomic cloud
\cite{PhysRevA.71.043807,QNDPolzik}.
\begin{figure}
\begin{center}
\includegraphics[width=0.8\columnwidth]{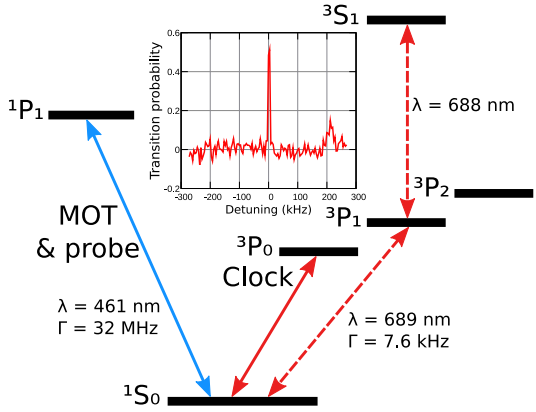}
\caption{Energy levels of Sr of interest here. The inset shows a
typical spectrum of the clock transition using the non-destructive
detection.} \label{fig:levels}
\end{center}
\end{figure}
\begin{figure}
\begin{center}
    \includegraphics[width=0.8\columnwidth]{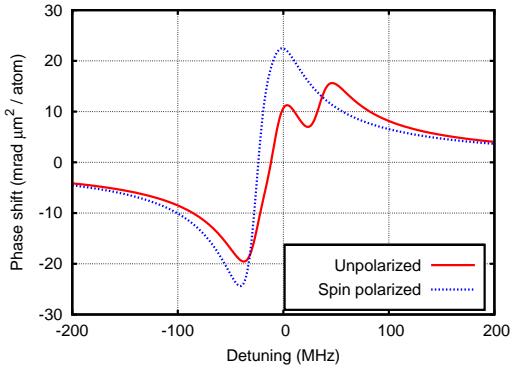}
\caption{Theoretical phase shift for the \blue\ transition with
zero magnetic field and a linearly polarized probe. It takes into
account the three different $F' = 7/2,\ 9/2$ and $11/2$ levels of
$^1\!P_1$, spanning over 60~MHz around their average frequency
(center of the plot). The phase shift is represented for equally
populated $m_F$ states (solid red curve) and spin-polarized atoms
in $m_F = 9/2$ or $m_F = -9/2$ states (dashed blue curve). For a
90~MHz detuning, these phase shifts are comparable and amount to a
few tens of mrad with a typical number of $N=10^4$ atoms.}
    \label{fig:signal}
\end{center}
\end{figure}
If the atomic resonance involves one of the two clock states,
the accumulated phase gives a measure of the number of atoms that
populate this state. When imposed after the clock interrogation,
the phase measurement can then yield the clock transition
probability.

\begin{figure}
\begin{center}
    \includegraphics[width=0.6\columnwidth]{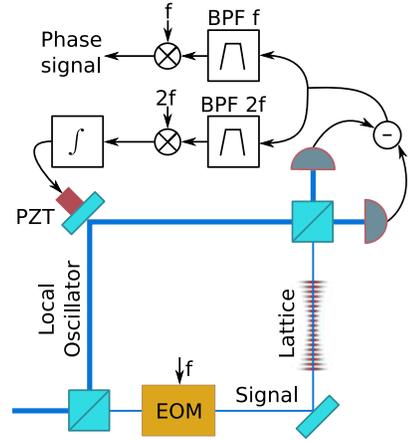}
\caption{Experimental setup. The number of atoms in the optical
lattice is proportionnal to the phase shift of the RF component at
the modulation frequency $f$, filtered by a band-pass filter
(BPF). The harmonic at frequency $2f$ is used to lock the phase of
the interferometer, hence maximizing the RF power of the signal
component.}
    \label{fig:setup}
\end{center}
\end{figure}

We have chosen to operate with the \blue\ transition (the relevant energy levels of Sr are plotted in Fig.
\ref{fig:levels}), for which the expected phase shift is plotted in
Fig.~\ref{fig:signal}. Two frequency
components detuned symmetrically around the resonance accumulate opposite phase shifts while passing through the atomic cloud. Their difference therefore gives a differential measure of the number of atoms. This is implemented using the first modulation sidebands induced by an electro-optic phase
modulator (EOM) in a Mach-Zender interferometer (MZI) as sketched in Fig.~\ref{fig:setup}.
A laser beam tuned to the \blue\ transition is split into a weak
signal and a strong local oscillator (LO). Their power is a few nW
and a couple of mW, respectively. The phase of the signal beam is
modulated at $f=90$~MHz by the EOM before traveling through the
atomic sample in the optical lattice. The electric field of the
signal beam is detected by a homodyne detection. The signal
interferes with the LO on the beam splitter closing the MZI and
the light intensities in each output arm of the beam splitter are
measured with fast Si photodiodes and electrically subtracted. The
output signal component at frequency $f$ is then demodulated,
giving a measure of the phase difference accumulated by the first
sidebands. It should be noted that this measurement is highly
differential, being immune to first order to the probe laser
frequency and amplitude noise, as well as to fluctuations of the
optical propagation lengths.

\subsection{Performance of the Detection Scheme \label{Sec:QND-Detection-noise}}
\begin{figure}
\begin{center}
    \includegraphics[width=\columnwidth]{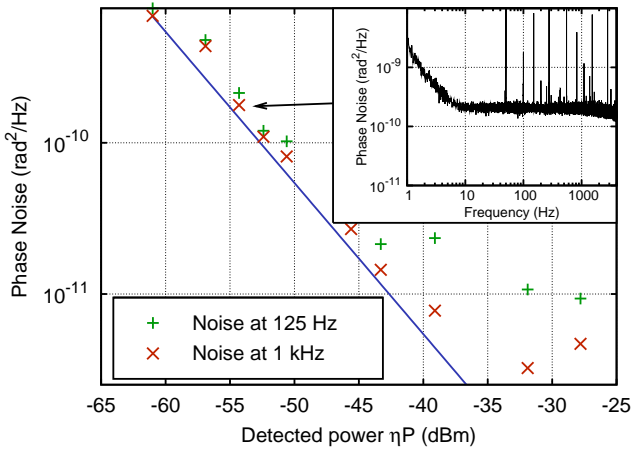}
    \caption{Detection noise power spectral density at 125~Hz and 1~kHz. The signal is shot noise limited (blue line) for powers up to 30~nW.
    The inset shows the full phase noise spectrum for a typical detected optical power $\eta P = 3$~nW, corresponding to a white noise level of $2\times 10^{-10}$~rad$^2$/Hz.}
    \label{fig:fft}
\end{center}
\end{figure}
The atomic population in $^1\!S_0$ is measured by applying two
consecutive probe pulses of duration $\tau_{nd}=3$~ms and typical
power $P = 12$~nW ($\eta P = 5$~nW, with $\eta$ being the
detection efficiency). The two pulses are separated by a 7~ms
interval during which the atoms are optically pumped in the dark
states $^3\!P_0$ and $^3\!P_2$ using the $^1\!S_0\,-\,^3\!P_1$ and
$^3\!P_1\,-\,^3\!S_1$ transitions (Fig.\,\ref{fig:levels}). The
second probe pulse does not experience the atomic phase shift and
then acts as a phase reference. A typical noise spectrum of the
phase signal is shown in Fig.~\ref{fig:fft}. It is shot noise
limited for Fourier frequencies higher than 10~Hz. The noise of
the resulting signal, as measured with no atoms in the lattice, is
0.4~mrad RMS for $\eta P = 5$~nW and scales as $1/\sqrt{P}$.

With about $N =10^4$ atoms in the lattice, the measured phase
shift is 40~mrad corresponding to a SNR of 100 per cycle, which is
close to the expected atomic quantum projection noise.

The measurement of the absolute transition probability associated
with the interrogation of the atomic ensemble with our clock laser
involves a third probe pulse to determine the total atom number.
The measured noise on the transition probability is
$\sigma_{\delta P}= 2$\% RMS with the previous parameters, varying
as $1/N$ for $N$ up to $10^4$. A typical spectrum of the clock
transition acquired with this method is shown in figure
\ref{fig:levels}.

Finally, a key aspect of the detection scheme performance is the
ability to recycle the atoms from one cycle to the other. The
fraction of atoms remaining in the lattice after the detection
pulses is measured to be larger than 0.95 for a lattice depth of
200~$E_R$.

\section{Optimization of the Strontium Clock Time Sequence\label{sec:SrStability}}

\begin{figure}
\begin{center}
    \includegraphics[width=0.8\columnwidth]{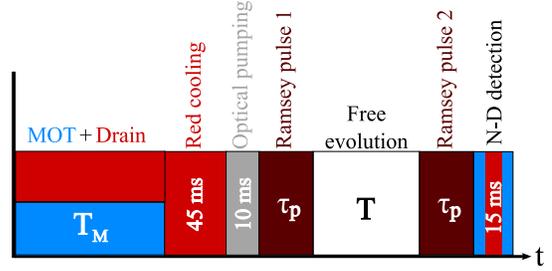}
    \caption{The time sequence for the Sr lattice clock. For further details, see \cite{targat:130801}.
    The minimum residual dead time $\widetilde{T}_d= T_d-T_M$ of this sequence is 70 ms.}
    \label{fig:SrSequence}
\end{center}
\end{figure}

The time sequence for operation of the Sr lattice clock is
sketched in Fig. \ref{fig:SrSequence}. The dead time $T_d$ can be
split up into two components, $T_d=T_M+\widetilde{T}_d$, where
$T_M$ is the capture time for the atoms, and $\widetilde{T}_d$ is
the time used for cooling, optical pumping, and detection of the
atoms. The present minimum residual dead time of the sequence is
$\widetilde{T}_d=70\,$ms, mainly limited by the duration of the
narrow line cooling in the lattice referred to as ``Red cooling''
on the figure. The duration of this cooling was adjusted so as to
optimize the atomic temperature in the lattice at a fixed laser
frequency and power. By allowing a variation of these parameters
the duration could certainly be shortened significantly. However,
to give a conservative estimate of the optimized clock stability
we keep this duration at its present value. The two parameters
left for optimization are therefore the duration of the capture
phase (``MOT+Drain'' on Fig. \ref{fig:SrSequence}) $T_M$ and the
Ramsey interrogation time $T$.

The optimal time sequence results from a balance between the Dick
effect and the detection noise. Taking both into account, the
Allan variance of the clock is given by
\begin{eqnarray}
\sigma_\text{tot}^2(\tau) &=& \sigma_y^2(\tau)
+\sigma_\text{det}^2(\tau), \label{eq:TotalAllan}
\end{eqnarray}
where $\sigma_y$ is defined in section \ref{sec:Dick} and
$\sigma_\text{det}$ is given by \cite{Santarelli98}
\begin{eqnarray}
\sigma_\text{det}(\tau) = \left( \frac{2}{\pi Q} \right)
\sigma_{\delta P} \sqrt{\frac{T_c}{\tau}}, \label{eq:Det-noise}
\end{eqnarray}
$Q$ and $\sigma_{\delta P}$ being the atomic quality factor and
the standard deviation of the detected transition probability.
$\sigma_{\delta P}$ scales as the inverse of the atom number $N$
up to $N=10^4$ for which  $\sigma_{\delta P}= 0.02$ as described
in section \ref{Sec:QND-Detection-noise}.

The non-destructive detection scheme allows recycling of the
atoms, so that the number of atoms after cycle $j$ is given by
\begin{eqnarray}
N_j &=& N_L + \xi N_{j-1} e^{-T_c/\tau_t},  \label{eq:CycleN}\\
\nonumber N_L &=& N_\text{max}
(1-e^{-T_M/\tau_t})e^{-(T_c-T_M)/\tau_t},
\end{eqnarray}
where $N_L$ is the number of atoms loaded into the optical lattice
in each cycle, $\tau_t$ is the lifetime of the cold atoms in the
lattice, and $N_\text{max}$ is, for a given $\tau_t$, the
maximally achievable number of atoms in the trap, that is for $T_M
\rightarrow \infty$. $\xi$ is the fraction of atoms kept in the
trap after a cycle. For our experiment, $N_\text{max} = \tau_t
\cdot 1.8 \cdot 10^4$/s, $\tau_t = 1.5 \;$s and $\xi = 0.95$.

From Eq. \eqref{eq:CycleN} we get the steady-state number of atoms
\begin{eqnarray}
\label{eq:NoA} N &=& N_\text{max} \frac{e^{T_M/\tau_t} -
1}{e^{T_c/\tau_t} - \xi},
\end{eqnarray}
% N &=& N_\text{max} \frac{e^{-(T_c-T_M)/\tau_t}-e^{-T_c/\tau_t
% }}{1-\xi e^{-T_c/\tau_t}}.
which can be used to find $\sigma_{\delta P}$ and hence
$\sigma_\text{det}(\tau)$ using Eq. \eqref{eq:Det-noise}, thus enabling
us to express $\sigma_\text{tot}(\tau)$ as a function of only
$T_M$ and $T$.

\begin{figure}
\begin{center}
    \includegraphics[width=1.0\columnwidth]{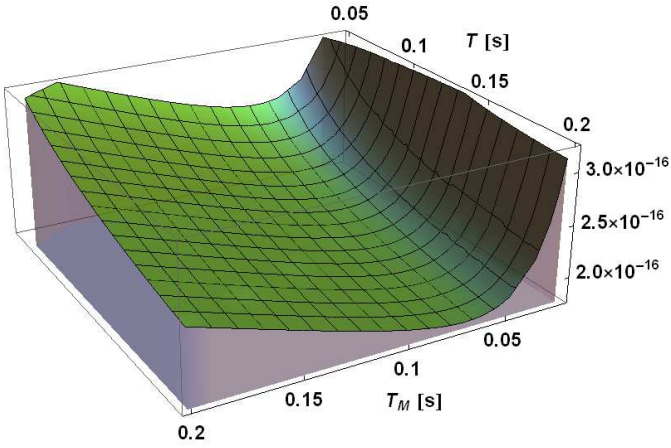}
    \caption{The total fractional Allan deviation at 1 s as a function of capturing time $T_M$
     and Ramsey dark time $T$ with residual dead time $\widetilde{T}_d=70 \,$ ms.}
    \label{fig:TotalAllan}
\end{center}
\end{figure}

Fig. \ref{fig:TotalAllan} displays
$\sigma_\text{tot}(1\,\textrm{s})$ as a function of both $T_M$ and
$T$. To remain in the validity domain of the model, we limited the
range of variation of $T$ up to 200\,ms as for
Fig.\,\ref{fig:FixedRatio}. Once again, the optimal $T$ is the
longest allowed one, $T=200\,$ms. The corresponding optimal value
for the loading time is $T_M=69\,$ms giving
$\sigma_\text{tot}(\tau)=1.8\cdot 10^{-16}\,\tau^{-1/2}$. The
individual contributions of the Dick effect and of the detection
noise are $\sigma_y(\tau) = 1.5 \cdot 10^{-16}\,\tau^{-1/2}$ and
$\sigma_\text{det}(\tau) = 1.0 \cdot 10^{-16}\,\tau^{-1/2}$,
respectively. Finally, the steady-state number of atoms in the
optimized configuration is $N=4000$.

\begin{figure}
\begin{center}
    \includegraphics[width=1.0\columnwidth]{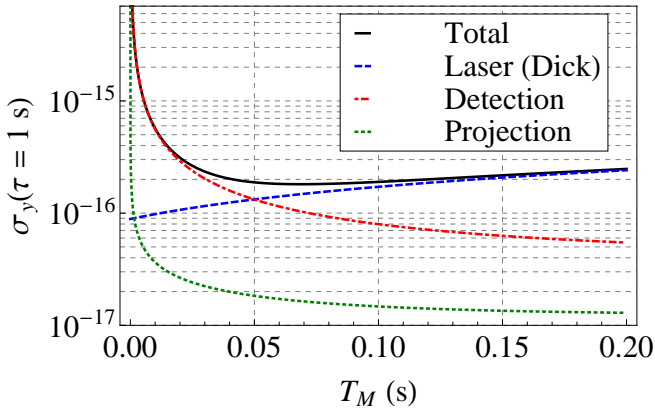}
    \caption{The different contributions to the total fractional Allan deviation at 1 s as a function of capturing time $T_M$
     for Ramsey dark time $T=200\,$ms with residual dead time $\widetilde{T}_d=70 \,$ ms.}
    \label{fig:NoiseContr}
\end{center}
\end{figure}

The individual contributions to $\sigma_\text{tot} (\tau=1\,$s)
for $T=200\,$ms are shown in Fig. \ref{fig:NoiseContr}. The
contribution from the quantum projection noise also is included in
the plot, showing that $\sigma_\text{tot}$ is still well above the
quantum limit, leaving room for further improvements. These
improvements would include increasing the trap lifetime and
reducing the residual dead time as well as enhancing the coherence time of the
interrogation laser.

\section{Conclusion \label{sec:Conclusion}}

We have shown that in parallel to the reduction of the
interrogation laser frequency noise, the optimization of the time
sequence could be a very efficient way to minimize the Dick effect
in optical lattice clocks. By using a non-destructive detection
scheme together with an adapted time sequence, the Allan deviation
of our clock could be optimized down to below $2\cdot
10^{-16}\,\tau^{-1/2}$, which would outperform current
state-of-the-art by about one order of magnitude.

Though very encouraging, this result is still about one order of
magnitude above the expected quantum limit of the clock. In the
optimized time sequence presented in section \ref{sec:SrStability}
the duty cycle is ``only'' $0.60$ and large room for improvement
remains. Cooling the atoms down to their minimal temperature
presently takes 45\,ms which could probably be strongly reduced by
using a more sophisticated time sequence, for instance allowing
both the frequency and power of the cooling laser to vary during
this phase. On the other hand, the lifetime of the atoms in the
lattice is presently 1.5\,s, so that about 20\,\% of
the atoms need to be reloaded at each cycle. This leads to a
relatively long loading time of 69\,ms in the optimized
configuration. We have not yet investigated in detail the limiting
factors of this lifetime in our setup but we see no fundamental
reasons preventing atoms from being kept in the lattice for ten
seconds or more. With such a lifetime, one would take full
advantage of the non-destructive detection scheme described in
section\,\ref{sec:Non-destr}, giving a $\sigma_\text{tot}
(\tau=1\,$s) on the order of $1\cdot 10^{-16}$ or below.

SYRTE is a member of IFRAF (Institut Francilien de Recherche sur
les Atomes Froids). This work has received funding from the
European Community's Seventh Framework Programme, ERA-NET Plus,
under Grant Agreement No. 217257, as well as from IFRAF, CNES and
ESA.

\bibliographystyle{IEEEtran}
\bibliography{QND-IEEE}

\end{document}